\documentclass[twocolumn,showpacs,preprintnumbers,amsmath,amssymb,prl]{revtex4}

\usepackage{graphicx}
\usepackage{dcolumn}
\usepackage{bm}

\begin{document}

\preprint{APS/123-QED}
\title{Diffusion thermopower of (Ga,Mn)As/GaAs tunnel junctions}
\author{Ts. Naydenova, P. D\"urrenfeld, K. Tavakoli, N. P$\acute{e}$gard, L. Ebel, K. Pappert, K. Brunner, C. Gould, L.W. Molenkamp}
 \affiliation{Physikalisches Institut (EP3), Universit\"at W\"urzburg,\\ Am Hubland,
D-97074 W\"urzburg, Germany}
\date{\today}
\begin{abstract}
We report the observation of tunneling anisotropic magnetothermopower, a voltage response to a temperature difference across an interface between a normal and a magnetic semiconductor. The resulting voltage is related to the energy derivative of the density of states in the magnetic material, and thus has a strongly anisotropic response to the direction of magnetization in the material. The effect will have relevance to the operation of semiconductor spintronic devices, and may indeed already play a role in correctly interpreting the details of some earlier spin injection studies.
\end{abstract}
\begin{flushright}
\end{flushright}
 
\maketitle

The most pressing issues of modern information technologies are power consumption and heat dissipation. As such, a proper understanding of the relationship between electrical and thermal effects in devices is essential to their design and operation. When these devices are magnetic in nature, the study of this relationship is the topic of the field of spin caloritronics \cite{Bauer2010}, which was strongly inspired by the initial reports of the spin Seebeck effect in metals \cite{Uchida2008} and subsequently in magnetic semiconductors \cite{Jaworski2010}. 

As a simple example of the importance of thermal effects on device operation, consider the problem of the injection of spin polarized current into a nonmagnetic semiconductor. Spin injection is often measured using a four contact nonlocal technique \cite{Jedema2001,Lou2007}, where current is passed between two adjacent contacts, and the chemical potential resulting from spin accumulation is detected at two further contacts which are situated next to, but not between the current contacts. Ideally, the signal from such measurements should produce a voltage at the detection leads, which is equal but opposite in sign depending on the relative orientations of the contacts. In practice this is almost never the case, as the measurements typically include a constant voltage offset, which it has become customary to neglect as an unimportant background contribution \cite{Lou2007}. 

This background voltage is a thermoelectric effect \cite{2010PhRvL.105m6601B}, which in most cases is isotropic to magnetic field and thus independent of the magnetization direction of the contacts. The subtraction of such a constant background is thus inconsequential to the experimental conclusions.

As we show in this Letter however, in materials where strong spin-orbit coupling links
the magnetic properties to the density of states \cite{Gould2007}, a spin caloritronic phenomenon, which we dub tunneling anisotropic magnetotermopower (TAMT), appears. This not only may have interesting implications for device applications, but can also lead to misinterpretations if one fails to take it into consideration. Indeed, we show that thermal voltages detected in such materials, e.g. the ferromagnetic semiconductor (Ga,Mn)As, closely resemble typical nonlocal spin-valve signals \cite{Ciorga2009}, even when no spin injector is used.

The configuration we study looks at the thermopower across a tunnel junction between ferromagnetic and nonmagnetic regions \cite{McCann2004}. Specificly, our experiment examines diffusion thermopower through a low-temperature grown i-GaAs tunnel barrier between a magnetic (Ga,Mn)As layer at 4.2~K and a highly Si doped
GaAs channel in which the electronic system is heated by an ac current. Previous
reports of thermopower in (Ga,Mn)As \cite{Pu2006} focused on the much larger phonon-drag
contribution which dominates thermopower in the bulk of the material. By designing our structure with a p-n interface and applying the heat directly to the electron system \cite{Gallagher1990, Molenkamp1990, Molenkamp1992}, phonon-drag terms do not contribute to the measured thermovoltages \cite{Jong1995, MOLENKAMP1992a}. We can thus extract the effects of magnetic anisotropies on the diffusion thermopower term. This carrier-related thermopower probes the energy derivative of the (Ga,Mn)As density of states (DOS) and thus shows a magnetic anisotropy similar to (Ga,Mn)As tunneling anisotropic magnetoresistance TAMR \cite{Gould2004}. This is clearly distinct from the twofold anisotropic magnetoresistance AMR-like symmetry observed in Ref. \cite{Pu2006} for the phonon-drag dominated Seebeck coefficient.

\begin{figure}
\includegraphics[width=0.40\textwidth]{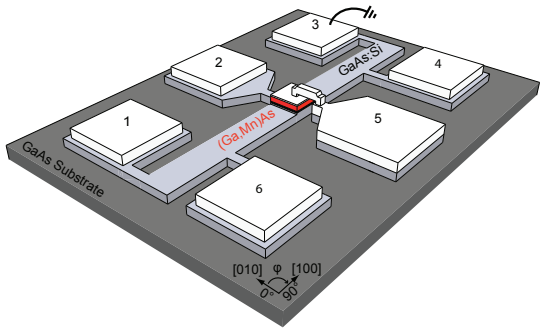}
\caption{\label{fig:schematic} Schematic of our TAMT device showing the geometry of all contacts as well as the crystal orientation. Heating current is passed through the Si-GaAs channel, and the thermovoltage is detected between contacts 2 and 5. The remaining contacts are used to inject the heating current and monitor the weak localization.}
\end{figure}

Our structure has a single magnetic (Ga,Mn)As contact
(20~nm thick), which is separated from a highly doped Si:GaAs mesa (60~nm) by a
1~nm low-temperature GaAs tunnel barrier. The semiconductor layers are grown on
a semi-insulating GaAs substrate and buffer by means of low-temperature
molecular beam epitaxy. Subsequently, and without breaking the vacuum, the
sample is transferred to a UHV electron-gun evaporation chamber where Ti/Au
contacts are deposited on the (Ga,Mn)As layer. The sample is patterned into the shape depicted in Fig.~\ref{fig:schematic} using a multistep optical lithography and
etching process. Contacts to the GaAs channel are established by Ti/Au
evaporation and lift-off. The $50~\times~50~\mu m^2$ pillar is connected to a
bonding pad through a gold air bridge fabricated using electron-beam
lithography and a multiple acceleration voltage process \cite{Borzenko2004}.

Transport measurements are performed at 4.2~K in a variable temperature
magnetocryostat fitted with a vector field magnet that allows for the application of a
magnetic field of up to 300~mT in any direction. As expected from our targeted electron heating method \cite{Jong1995, MOLENKAMP1992a}, the (Ga,Mn)As contact remains at 4.2~K throughout the experiment, as confirmed by verifying that the anisotropy
fingerprints of the (Ga,Mn)As remains unchanged when applying a heating current, and comparing that to the known temperature dependence of such fingerprints \cite{Pappert2007a}.

The electron system in the GaAs channel is heated out of equilibrium with respect to the GaAs lattice \cite{Jong1995, MOLENKAMP1992a} by a 13~Hz ac current applied across the channel, with one of its contacts defining the system ground. The resulting temperature difference $\Delta T$ across the tunnel barrier generates a thermal voltage $V_{th}$ \cite{Barnard1972} which can be detected between contacts 2 and 5, as shown in Fig.~\ref{fig:schematic}. The amplitude of this thermal voltage depends on $\Delta T$, or equivalently on the electron temperature in the GaAs channel. Since we heat at sufficiently low frequencies that the electron temperature remains in quasistatic equilibrium and follows the applied power, the thermal voltage will be related to the square of the applied voltage, and can be detected at the double frequency 2$f$ using standard lock-in amplifier techniques.

To establish a relationship between the temperature of the carrier system in
the GaAs channel and the heating current, we record the magnetoresistance of the
GaAs channel in response to an out-of-plane magnetic field for various applied voltages. The width of the weak localization feature \cite{Bergmann1984} as a function of heating current is then compared to the width of the same feature taken at varying cryostat temperature using low current. Doing so, we find that a heating areal current density of 5.5 A/m through the heating channel corresponds to an electron temperature of $\approx$ 22 K, to an estimated accuracy of about $\pm 1.5$ K.

\begin{figure}[h]
   \centering
       \includegraphics[width=0.40\textwidth]{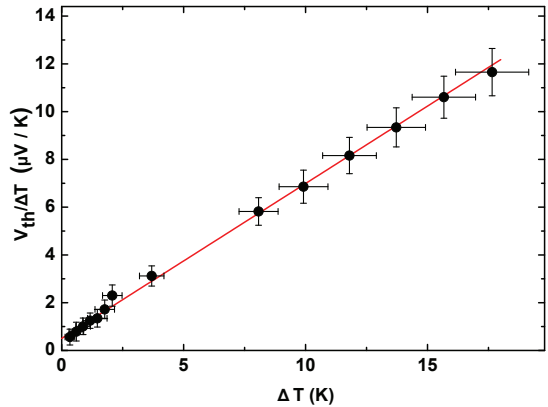}
   \caption{Amplitude of the thermopower signal as a function of the temperature in the heating channel. A magnetic field of 300 mT is applied along the [100] crystal direction.}
   \label{fig:Tdep}
\end{figure}

Fig.~\ref{fig:Tdep} shows the measured thermopower $\zeta = \frac{V_{th}}{\Delta
T}$ for different thermal gradients $\Delta T$ recorded at a field of 300~mT
along the [100] magnetic easy axis of the (Ga,Mn)As layer. Similar values are
obtained for other magnetic field directions and amplitudes. The measured voltage is clearly temperature dependent and has a magnitude in the $\mu$V range, comparable to the background voltages typically observed in nonlocal measurements.

The intercept with the y-axis S (4.2~K) $= \lim_{\Delta T \to 0} \zeta = 
\frac{V_{th}}{\Delta T}$ yields a Seebeck coefficient of \textbf{0.5}
$\mu$VK$^{-1}$. This number represents the diffusion part of the Seebeck
coefficient at 4.2~K and is 3 orders of magnitude smaller than the phonon-drag
dominated Seebeck coefficient measured in (Ga,Mn)As by Pu et al. \cite{Pu2006}.

This value is consistent with an estimate of the diffusion Seebeck coefficient at 4.2~K
from a simple model of detailed balance across a p-n interface under the assumption that the carrier distribution on each side of the interface is represented by a Fermi function at a different temperature \cite{Esaki1958}. We further assume that the thin tunnel barrier at the interface in our device can be approximated to be large compared the transport energy scales and that the tunnel probability is energy independent. We thus model it by a delta function.

This leads to an expression for total current through the interface given by:

\begin{eqnarray}
\mathbf{j_{total}} = A \int_{E_{C,GaAs:Si}}^{E_{V,(Ga,Mn)As}-eV_{th}}F(E)dE
\end{eqnarray}
where
\begin{eqnarray}
 F(E)=D_{GaAs:Si}(E)\cdot D_{(Ga,Mn)As}(E-eV_{th})\times \nonumber\\ (f_{GaAs:Si}(E)-f_{(Ga,Mn)As}(E-eV_{th})).
 \label{eq:Thermostrom3}
 \end{eqnarray}

Here, $D_{GaAs:Si}(E)$ and $D_{(Ga,Mn)As}(E-eV_{th})$ are the GaAs and (Ga,Mn)As density of states, and $f_{GaAs:Si}(E)$ and $f_{(Ga,Mn)As}(E-eV_{th})$ are the Fermi functions in the two regions. \textit{A} is a constant which includes geometry terms as well as the tunneling probability, and which is of no importance to the analysis. Since the total current flowing must be zero at equilibrium, given the density of states of the two materials the above equation can be solved numerically for $V_{th}$. 

For an order of magnitude estimate we assume a simple parabolic band model. The carrier density in the GaAs layer is $n = 4 \cdot 10^{18} cm^{-3}$ (from Hall measurements) and we estimate a value of $p = 1 \cdot 10^{21}  cm^{-3}$ for the (Ga,Mn)As (typical for layers with similar growth conditions). The effective mass in GaAs is known and for the (Ga,Mn)As, an average hole mass of 0.5$m_e$ is used. This calculation leads to a value of 0.3 $\mu VK^{-1}$ for the diffusion Seebeck coefficient, consistent with the experimentally obtained value of 0.5 $\mu VK^{-1}$

This model also predicts the linear dependence of $\zeta$ on temperature as expected in general for the diffusion Seebeck effect \cite{Cutler1969} and is in qualitative agreement with Fig.~\ref{fig:Tdep}. While a quantitative model would require a detailed treatment of the top four valence band levels, including their anisotropic effective mass, the simplified degenerate semiconductor model implying a Fermi level deep in the (Ga,Mn)As valence band is sufficient to capture the essence of the results.

    While this linear dependence is expected from the diffusive contribution to thermopower, phonon-drag contributions are typically highly nonlinear and usually even nonmonotonic \cite{Barnard1972}. Our observation of linear behavior is therefore further evidence that we are measuring the intrinsic carrier thermopower.

\begin{figure}[h]
   \centering
       \includegraphics[width=0.4\textwidth]{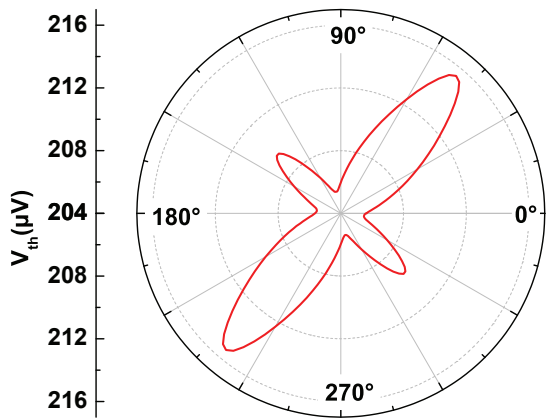}
       \caption{Thermovoltage in a magnetic field of 300~mT as a function of field angle. $0^{\circ}$ is along [010] crystal
direction.}
   \label{fig:VthPolar}
\end{figure}

    Since the (Ga,Mn)As DOS is the only magnetic field dependent term in Eq.~\ref{eq:Thermostrom3}, thermopower measurements reflect its response to a magnetic field. More specifically, the product of $D_{(Ga,Mn)As}(E)$ with a Fermi function leads to a response which is proportional to the energy derivative of the DOS of (Ga,Mn)As at the Fermi level. This implies that the observed signal should reflect the properties of the magnetic anisotropy of the (Ga,Mn)As DOS.
    
    Figure~\ref{fig:VthPolar} shows the measured thermovoltage in our
    sample versus the direction of a 300 mT magnetic field rotated in the sample
    plane and thus continuously perpendicular to the temperature gradient. 
    The magnetic field angle is given with
    respect to the [010] crystal direction as denoted in
    Fig.~\ref{fig:schematic}. We observe four equivalent minima close
    to the [100] and [010] crystal axes and two sets of local maxima: a
    larger one along [110] and a smaller one along $[\overline{1}10]$. This symmetry is strongly reminiscent of (Ga,Mn)As TAMR measurements which map the symmetry of the (Ga,Mn)As DOS \cite{Gould2004} and is consistent with the
    thermovoltage mapping the energy derivative of the (Ga,Mn)As DOS as it depends on magnetization direction.

    Because it maps the derivative of the (Ga,Mn)As DOS, the symmetry of the
    diffusion thermopower term in Fig.~\ref{fig:VthPolar} is very
    distinct from the AMR-like symmetry \cite{McGuire1975,Baxter2002} 
    observed in Ref. \cite{Pu2006} for the phonon-drag dominated Seebeck
    coefficient. AMR has a cos$^{2}\theta$ dependence on the angle $\theta$ between current
     and magnetization and thus a twofold symmetry with a minimum in resistance when the
     magnetization points along the current direction and a 
     maximum for magnetization perpendicular
     to the current.
    In our sample, the current in the GaAs heating channel is along the
    90$^\circ$  direction, while the tunnel interface between GaAs
    and (Ga,Mn)As is not in the current path. The location of
    minima and maxima in Fig.~\ref{fig:VthPolar} is crystal direction
    dependent and has no relation to the current direction, as
    confirmed on an identical structure with the heating channel
    oriented along [100], and thus rotated 90 degrees compared to
    Fig.~\ref{fig:schematic}.

\begin{figure}[h]
   \centering
       \includegraphics[width=0.5\textwidth]{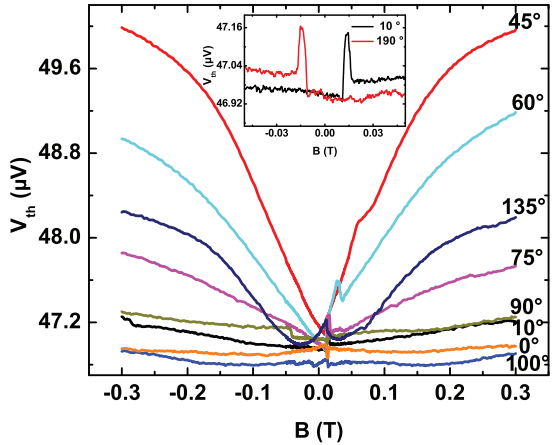}
       \caption{Thermovoltage as a function of magnetic field swept along various in-plane directions. Inset: Enlargement of the low field part. The 190$^\circ$ curve is offset by 0.15 $\mu V$ to remove a small thermal drift.}
   \label{fig:VthB}
\end{figure}
    For a more detailed investigation of the magnetic anisotropy, we
    record the thermopower during high resolution magnetic field scans
    from -300~mT to +300~mT along many in-plane crystal directions
    (Fig.~\ref{fig:VthB}). The electron temperature of the GaAs heating channel
    is kept constant at 12 K during the whole measurement, while the lattice temperature in the whole sample, as well as the carrier population in the (Ga,Mn)As layer remain at 4.2~K. The recorded thermovoltage is a
    function of the magnetization direction in the (Ga,Mn)As layer and
    can again be explained by considering the (Ga,Mn)As magnetic anisotropy.

    When the magnetic field is swept from -300~mT to 300~mT,
    the magnetization rotates towards an easy axis of the material,
    here towards  $0^\circ$  and  90$^\circ$  as the field approaches
    0~mT. It reverses at small positive fields in a mostly two-jump
    switching event. As the magnetization rotates, the
    thermovoltage changes smoothly. The largest magnetization rotation
    angle and thus largest change in thermovoltage is observed when
    the magnetic field is swept along the 45$^\circ$  hard axes of the
    (Ga,Mn)As material ([110]). When sweeping the field along or close to
    a magnetic easy axis ( 0$^\circ$ or  90$^\circ$), the magnetization
    stays along this easy axis direction, thus the thermovoltage is
    almost constant throughout the sweep apart from the one- or
    two-jump switching event. The figure thus shows two distinct components of TAMT: a volatile contribution caused by the field holding the magnetization away from its easy axes at high fields and a nonvolatile contribution associated with the difference in density of states when the magnetization switches between the two easy axis.

\begin{figure}
        \includegraphics[width=0.4\textwidth]{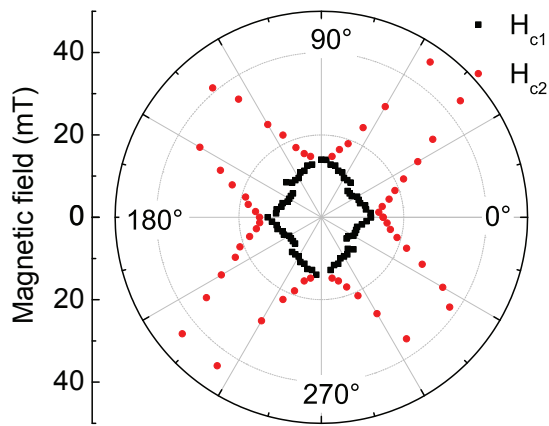}
        \caption{Polar plot of the magnetic field positions of the switching events H$_{c1}$ and H$_{c2}$ in the thermopower signal.}
   \label{fig:Hc1Hc2polar}

\end{figure}

    It is worth noting that the switching features in the thermovoltage curves in Fig.~\ref{fig:VthB} could easily be
    mistaken for the detection of a nonlocal spin signal. As an
    example, the inset displays the $190^\circ$ curve (field scanned from
    left to right) together with the $10^\circ$ curve (field scanned from
    right to left) in spin-valve fashion showing a similar spin-valve signal as reported in Ref.\cite{Ciorga2009}.
    
    We extract the first and second switching fields of the magnetization reversal, H$_{c1}$ and H$_{c2}$ respectively, and plot them radially as a function of the field orientation
    $\varphi$ in Fig.~\ref{fig:Hc1Hc2polar}. This plot reveals the
    (Ga,Mn)As anisotropy components, just as in previous crystalline AMR \cite{Pappert2007} and TAMR \cite{Gould2004} measurements. The
    global easy axes correspond to the diagonals of the rectangular
    pattern of the first switching fields H$_{c1}$. As typical for
    (Ga,Mn)As \cite{Gould2008}, two small uniaxial anisotropy
    components along the [110] (elongation of the center square) and
    along the [100] (step features in the central shape) crystal
    direction are visible in addition to the main biaxial anisotropy.

    The behavior of the measured voltage is thus fully
    described by the magnetization direction in the (Ga,Mn)As layer, together with the
    thermovoltage dependence on the magnetization direction given in Fig.~\ref{fig:VthPolar}.
    
    In summary, diffusion thermopower can be measured over a tunnel
    barrier between a magnetic (Ga,Mn)As layer at 4.2~K and a heated
    carrier system in a GaAs mesa. This diffusive component of the Seebeck coefficient is in
    accordance with theoretical estimates, and much smaller than the
    phonon-drag dominated Seebeck coefficient reported by Pu et al.
    \cite{Pu2006}. Its magnetic field dependence is given by the
    derivative of the (Ga,Mn)As DOS and is also clearly different from
    the AMR-like symmetry observed in Ref. \cite{Pu2006}. Moreover, this dependence of the thermopower on derivative of DOS could prove a useful tool in mapping out the details of the DOS, especially in samples near the metal-insulator transition.

    \vspace{0.1cm}
    \begin{acknowledgments}
    The authors thank G. Schmidt, S. Mark, and J. Constantino for useful
    discussions. We acknowledge financial support by the EU (Nanospin and SemiSpinNet), and the German DFG (BR 1960/
4-1).
    \end{acknowledgments}
    
    \textit{Note added} - During the review of this Letter, three related papers \cite{Walter2011,LeBreton2011,Liebing2011} appeared on spin thermopower across tunnel junction. These relate to metallic systems, and do not show anisotropic behavior.


\begin{thebibliography}{10}

\bibitem{Bauer2010}
G.~E. Bauer, A.~H. MacDonald, and S. Maekawa, Solid State Communications {\bf
  150},  459  (2010).

\bibitem{Uchida2008}
K. Uchida {\it et~al.}, Nature {\bf 455},  778  (2008).

\bibitem{Jaworski2010}
C.~M. Jaworski {\it et~al.}, Nat Mater {\bf 9},  898  (2010).

\bibitem{Jedema2001}
F.~J. Jedema, A.~T. Filip, and B.~J. van Wees, Nature {\bf 410},  345  (2001).

\bibitem{Lou2007}
X. Lou {\it et~al.}, Nat Phys {\bf 3},  197  (2007).

\bibitem{2010PhRvL.105m6601B}
F.~L. Bakker, A. Slachter, J.-P. Adam, and B.~J. van Wees, Physical Review
  Letters {\bf 105},  136601  (2010).

\bibitem{Gould2007}
C. Gould, K. Pappert, G. Schmidt, and L. Molenkamp, Adv. Mater. {\bf 19},  323
  (2007).

\bibitem{Ciorga2009}
M. Ciorga {\it et~al.}, Phys. Rev. B {\bf 79},  165321  (2009).

\bibitem{McCann2004}
E. McCann and V.~I. Fal'ko, Journal of Magnetism and Magnetic Materials {\bf
  268},  123   (2004).

\bibitem{Pu2006}
Y. Pu, E. Johnston-Halperin, D.~D. Awschalom, and J. Shi, Phys. Rev. Lett. {\bf
  97},  036601  (2006).

\bibitem{Gallagher1990}
B.~L. Gallagher {\it et~al.}, Phys. Rev. Lett. {\bf 64},  2058  (1990).

\bibitem{Molenkamp1990}
L.~W. Molenkamp {\it et~al.}, Phys. Rev. Lett. {\bf 65},  1052  (1990).

\bibitem{Molenkamp1992}
L.~W. Molenkamp {\it et~al.}, Phys. Rev. Lett. {\bf 68},  3765  (1992).

\bibitem{Jong1995}
M.~J.~M. de~Jong and L.~W. Molenkamp, Phys. Rev. B {\bf 51},  13389  (1995).

\bibitem{MOLENKAMP1992a}
L.~W. Molenkamp, M.~J.~P. Brugmans, H. van Houten, and C.~T. Foxon,
  Semiconductor Science and Technology {\bf 7},  B228  (1992), and references therein.

\bibitem{Gould2004}
C. Gould {\it et~al.}, Physical Review Letters {\bf 93},  117203  (2004).

\bibitem{Borzenko2004}
T. Borzenko, C. Gould, G. Schmidt, and L.~W. Molenkamp, Microelectronic
  Engineering {\bf 75},  210  (2004).

\bibitem{Pappert2007a}
K. Pappert {\it et~al.}, New Journal of Physics {\bf 9},  354  (2007).

\bibitem{Barnard1972}
R.~D. Barnard, {\em Thermoelectricity in Metals and Alloys} (Taylor and Francis
  LTD., London, 1972).

\bibitem{Bergmann1984}
G. Bergmann, Physics Reports {\bf 107},  1  (1984).

\bibitem{Esaki1958}
L. Esaki, Phys. Rev. {\bf 109},  603  (1958).

\bibitem{Cutler1969}
M. Cutler and N.~F. Mott, Phys. Rev. {\bf 181},  1336 (1969).

\bibitem{McGuire1975}
T.~R. McGuire and R.~I. Potter, IEEE Trans. Magn., 11, 1018 (1975).

\bibitem{Baxter2002}
D.~V. Baxter {\it et~al.}, Physical Review B {\bf 65},  212407  (2002).

\bibitem{Pappert2007}
K. Pappert {\it et~al.}, Applied Physics Letters {\bf 90},  062109  (2007).

\bibitem{Gould2008}
C. Gould {\it et~al.}, New Journal of Physics {\bf 10},  055007  (2008).

\bibitem{Walter2011}
M. Walter {\it et~al.}, Nat Mater., 10, 742 (2011).

\bibitem{LeBreton2011}
J.-C. Le~Breton {\it et~al.}, Nature {\bf 475},  82  (2011).

\bibitem{Liebing2011}
N. Liebing {\it et~al.}, Phys. Rev. Lett. {\bf 107},  177201  (2011).

\end{thebibliography}

\end{document}